\documentclass{aastex6}

\newcommand\bc{$\beta$ CMi}
\newcommand\be{Be}

\begin{document}

\title{A Spectroscopic Orbit for the late-type \be\ star \bc}


\author{Nick Dulaney, Noel D. Richardson\altaffilmark{1}, Cody J. Gerhartz, J. E. Bjorkman, K. S. Bjorkman,}
\affil{Ritter Observatory, Department of Physics and Astronomy, The University of Toledo, Toledo, OH 43606-3390, USA}

\author{Alex C. Carciofi}
\affil{Instituto de Astronomia, Geof\'isica e Ci\^encias Atmosf\'ericas, Universidade de S\~ao Paulo, SP 05508-900, Brazil}

\author{Robert Klement\altaffilmark{2},}
\affil{European Southern Observatory, Alonso de C\'ordova 3107, Vitacura, Casilla 19001, Santiago, Chile}

\author{Luqian Wang,}
\affil{Center for High Angular Resolution Astronomy and Department of Physics and Astronomy, Georgia State University, P.O. Box 5060, Atlanta, GA 30302-5060, USA}

\author{Nancy D. Morrison, Allison D. Bratcher, Jennifer J. Greco, Kevin K. Hardegree-Ullman, Ludwik Lembryk, Wayne L. Oswald, Jesica L. Trucks}
\affil{Ritter Observatory, Department of Physics and Astronomy, The University of Toledo, Toledo, OH 43606-3390, USA}


\altaffiltext{1}{noel.richardson@UToledo.edu}
\altaffiltext{2}{Astronomical Institute of Charles University, Charles University, V Hole\v{s}ovi\v{c}k\'ach 2, 180 00, Prague 8, Czech Republic}

\begin{abstract}

The late-type Be star \bc\ is remarkably stable compared to other Be stars that have been studied. This has led to a realistic model of the outflowing Be disk by Klement et al. These results showed that the disk is likely truncated at a finite radius from the star, which Klement et al.~suggest is evidence for an unseen binary companion in orbit. Here we report on an analysis of the Ritter Observatory spectroscopic archive of \bc\ to search for evidence of the elusive companion. We detect periodic Doppler shifts in the wings of the H$\alpha$ line with a period of 170 d and an amplitude of 2.25 km s$^{-1}$, consistent with a low-mass binary companion ($M\approx 0.42 M_\odot$).
We then compared the small changes in the violet-to-red peak height changes ($V/R$) with the orbital motion. We find weak evidence that it does follow the orbital motion, as suggested by recent Be binary models by Panoglou et al. 
Our results, which are similar to those for several other Be stars, suggest that \bc\ may be a product of binary evolution where Roche lobe overflow has spun up the current Be star, likely leaving a hot subdwarf or white dwarf in orbit around the star. Unfortunately, no direct sign of this companion star is found in the very limited archive of {\it International Ultraviolet Explorer} spectra.

\end{abstract}


\section{Introduction} \label{sec:intro}

Classical Be stars are the best-studied example of rapidly rotating massive stars. These stars are non-supergiant B stars that show Balmer emission at least some of the time. Through yet undetermined processes, the star forms a thin gaseous equatorial decretion disk that rotates in a Keplerian manner (Kraus et al.~2012, Wheelwright et al.~2012). This disk is the source of the emission lines, produces an infrared excess, and can polarize the starlight by a small amount. 
The processes leading to the Be phenomenon are not understood, but there seems to be no significant changes in the frequency of Be occurrence from inferred ages of clusters containing Be stars (e.g., Abt 1979; McSwain et al.~2005).

\bc\ (B8Ve) was recently modeled by Klement et al.~(2015). The star was used to develop and fully test the viscous decretion disk models (Lee et al.~1991; Saio 1991; Carciofi et al. 2006) because of its relatively low amount of variability. A detailed history of the work on the star is included in Klement et al.~(2015), and we outline the most important observational results here. 
The star is located about 50 pc away, with no measurable reddening towards the star so that the measured colors agree well with expectations for the spectral type.
The disk was best modeled to have an inclination of $43^{+3^\circ}_{-2^\circ}$, with a position angle on the sky of $133^{+4^\circ}_{-3^\circ}$, measured from north to east. The star is rotating very near the critical rate ($W \gtrsim 0.98$). The star was suggested to be a binary by Jarad et al.~(1989), but the results had not been confirmed. More recently, Folsom et al.~(2016) presented an analysis of weak $V/R$ ratio variations (i.e., the ratio of the violet and red peak heights in a double-peaked emission line) with a period of 182.83 d.

Klement et al.~(2015) found that to match the long-wavelength (radio) portion of the SED, the decretion disk surrounding the Be star must be truncated at a relatively small radius from the star ($35^{+10}_{-5} R_{eq}$). Theoretical expectations for Be disks predict that the disk should flow outward with a slowly increasing radial velocity and decreasing density all the way out to the sonic point in the disk, which is defined as where the outflow velocity equals the local speed of sound (Rivinius et al. 2013). This point is typically at a distance of few 100 stellar radii, much larger than the disk in the \bc\ system. In a binary system, hydrodynamical simulations show that the companion star truncates the disk with a steep fall-off in density somewhere between the 3:1 and 2:1 resonance radius (the exact locations depends on the viscosity parameter and on the binary mass ratio; Okazaki et al.~2002; Panoglou et al.~2016). 

There were three additional indications of the binary nature of \bc\ from the analysis of Klement et al.~(2015). First, the calcium triplet (Ca II $\lambda8498, 8542, 8662$\AA) was found to be in emission with stronger emission than the hydrogen Paschen series. Polidan (1976) suggested that such emission was linked to binarity of Be stars. The Ca II $\lambda 3934$ line has an emission component narrower than the photospheric absorption. These Ca II emission features all point towards an origin in the outer disk, rather than the rapidly-rotating inner disk. Lastly, the ultraviolet C IV $\lambda 1548$ resonance line is present, showing signs of a wind moving at $\sim230$ km s$^{-1}$. This is not expected for such a late-type B star, and the velocity is quite slow, suggesting it may be from interactions between a hot secondary and the outer disk.

In addition to truncating the disk, the binary companion can perturb the disk density structure. For example, Panoglou et al.~(2016) showed how the companion can create standing density waves in the disk that are phase-locked to the binary, when the orbit is circular or of moderate eccentricity. Consequently, the weak $V/R$ variations with a 183 d timescale reported by Folsom et al.~(2016) are another hint of the possible presence of the binary companion. 

Given the large amount of circumstantial evidence pointing to a binary companion around \bc, it is important to attempt a firm detection of this companion star or its orbital motion. This paper aims to examine the properties of the Be star \bc\ in the context of disk truncation and standing density waves caused by a binary companion. We have observed \bc\ since the year 2000 with high resolution optical spectroscopy. The observations are presented in \S 2. In \S 3, we present our single-lined spectroscopic orbit of the system. In \S 4, we search for variations in the density of the disk relative to the binary orbit. We discuss these findings in \S 5 and conclude our study in \S 6.

\section{Observations} \label{sec:obs}

We collected 124 high-resolution spectra of \bc\ between 2000 January 28 and 2016 May 6 with the 1.06-m Ritter Observatory telescope and a fiber fed \'echelle spectrograph. The \'echelle spectrograph produces spectra with a resolving power of $R = 26,000$. Prior to 2007 April, the spectra were recorded on a $800\times1200$ pixel EEV CCD05--20--0--202 with 22.5$\mu$m pixels. The resulting spectra included nine non-adjacent \'echelle orders spanning 5280\AA\ to 6600\AA. The detector was upgraded in 2007 April to a Spectral Instruments 600 Series camera, with a $4096\times 4096$ pixel CCD with 15$\mu$m pixels. The new detector expanded the spectral range of all \'echelle orders, and extended the total coverage to span from $\sim 4300 - 7000$\AA\ across 21 orders, with some order overlap in the blue. Unfortunately, the blue range of the spectra are fairly noisy due to the combination of fiber losses and instrumental response. 

In general, one spectrum was obtained per night, but occasionally several spectra were obtained in a single night. We treated the consecutive observations as independent measurements rather than co-adding the nightly spectra, and the agreement in measurements was within the formal errors of the measurements. The order containing H$\alpha$ typically had the best quality of the recorded spectrum, with a $S/N\approx 75-120$ per pixel. The data were all reduced with standard IRAF\footnote{IRAF is distributed by the National Optical Astronomy Observatory, which is operated by the Association of Universities for Research in Astronomy (AURA) under a cooperative agreement with the National Science Foundation.} procedures to remove bias frames and flat field the \'echelle apertures. Wavelength calibration was accomplished with exposures of a ThAr lamp, with typical residuals of the fit on the order of $<0.01$ \AA\ ($< 0.4$ km s$^{-1}$). The H$\alpha$ region was telluric corrected by means of template spectra created with observations of rapidly rotating hot stars from the same instrument, and then adjusted interactively with the IRAF task {\tt telluric}.


\section{A Single-Lined Orbit for \bc } \label{sec:orbit}

Our optical spectra are of reasonably good quality ($S/N\approx 75-100$) in the red portion of the spectrum. In the wavelength regions where the better signal may allow us to measure weak photospheric absorption lines for radial velocities, the best lines are the \ion{Si}{2} $\lambda\lambda$6347,6371 absorption lines. Unfortunately, these lines are very weak in late-type main-sequence B stars, and the rotational broadening of these lines ($v \sin i = 270$ km s$^{-1}$; Klement et al.~2015) coupled with the depth of a few percent of the continuum level, makes it difficult to measure accurately the radial velocities in spectra with $S/N\sim100$. Errors on measurements of these lines were on the order of 10--15 km s$^{-1}$.
The deeper photospheric absorptions occur in the blue parts of the spectrum (e.g., \ion{He}{1} 4471, \ion{Mg}{2} 4481) which typically have a $S/N\lessapprox 50$, making radial velocity measurements difficult to extract either via cross correlation or through Gaussian fits. H$\beta$ has a similar low quality, with emission from the Be disk contaminating the core of the absorption. 

\begin{figure} 
\begin{center} 
\includegraphics[angle=90, width=8cm]{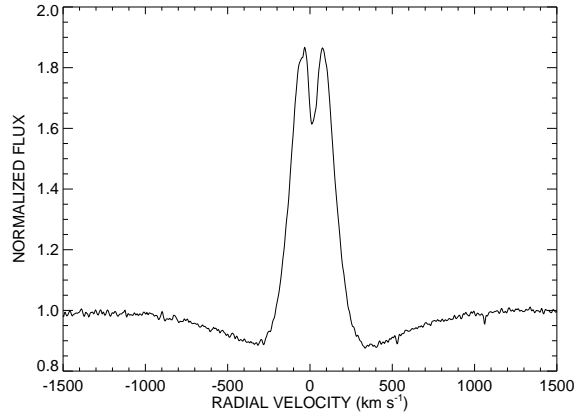}

\end{center} 
\caption{A typical H$\alpha$ spectrum of \bc\ used in this study. The edges of the plot roughly correspond to the edge of the spectrum with the old camera. } 
\label{spec} 
\end{figure}

For these reasons, the best line to study in this data set is the H$\alpha$ line. The line shows photospheric absorption wings at high velocities and a double-peaked emission in the line core, as shown in the example spectrum in Fig.~\ref{spec}. Note that although the older camera setup does not allow us to fully reach the continuum in the 90\AA\ spectral window, the flux should be within a percent of the continuum at high velocities, based on our comparisons to the photospheric model presented in Klement et al.~(2015). We measured the bisectors of the line wings using the cross correlation method described by Shafter et al.~(1986). This method has successfully been applied in multiple studies of Be binaries to find small companions, such as $\gamma$ Cas (Smith et al.~2012). The benefit of this technique is that it samples the fastest moving gas in the disk emission, which is the gas closest to the star in a Keplerian rotating disk, and should represent motion of the Be star and not structures in the disk. 

\begin{figure} 
\begin{center} 
\includegraphics[angle=90, width=8cm]{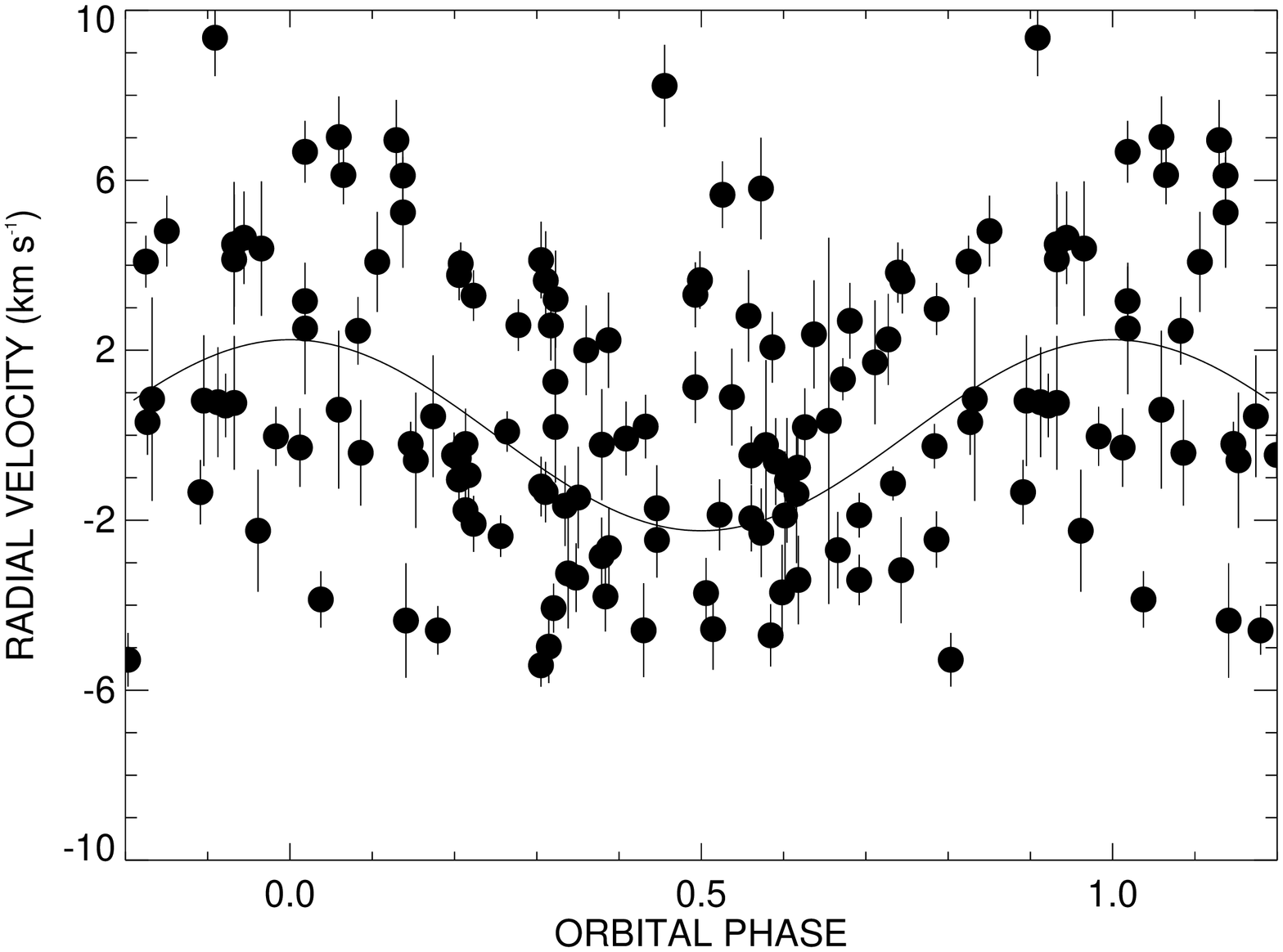}
\includegraphics[angle=90, width=8cm]{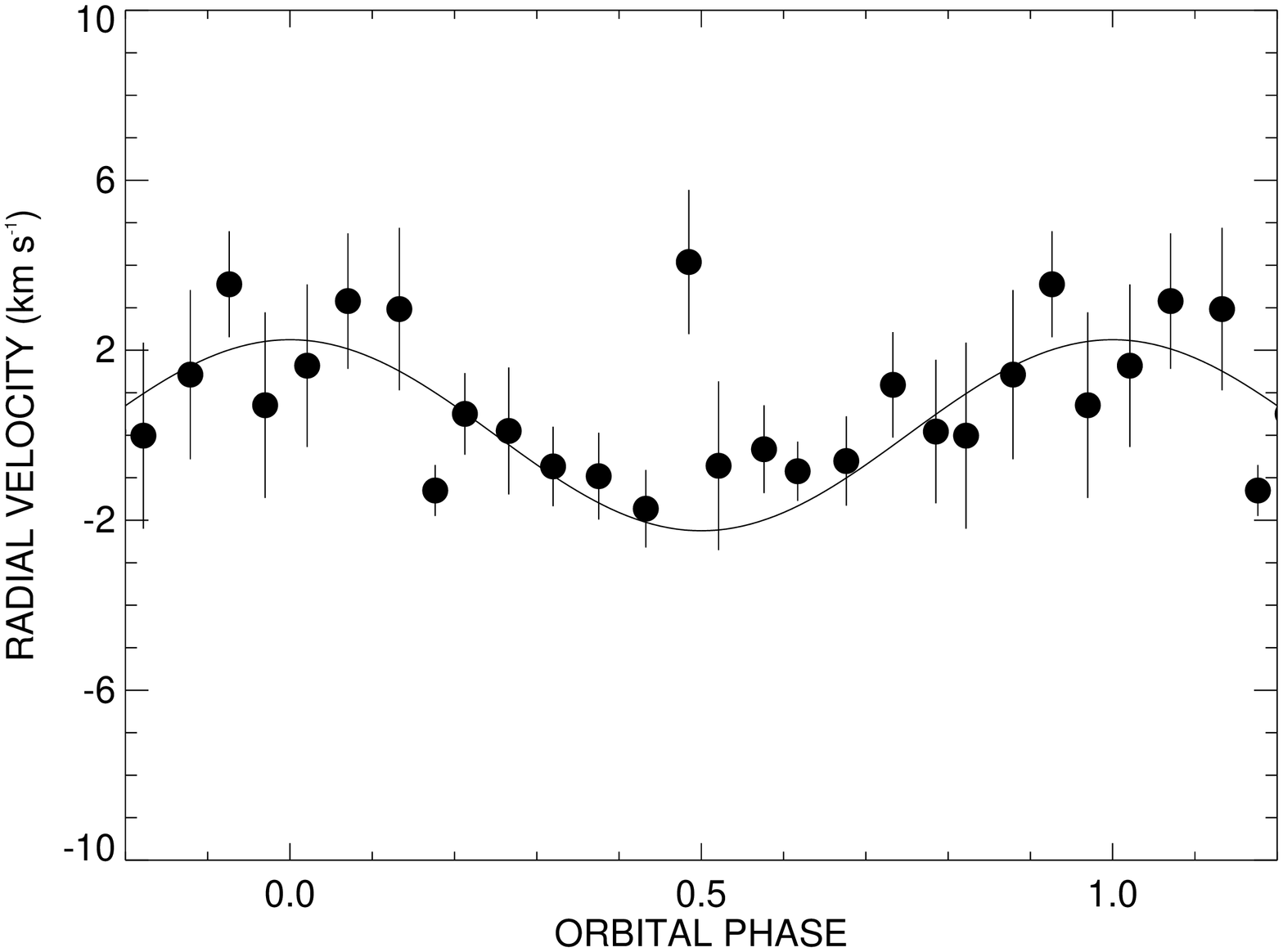}

\end{center} 
\caption{The orbital radial velocity curve (solid line) based upon the observed emission wing velocities (left), and a phase-binned radial velocity curve of the same data (right). The errors in the binned radial velocity curve represent the error of the mean. } 
\label{fig1} 
\end{figure}

The differential radial velocities (Table 1) were measured with errors of $\sigma \lesssim 0.7$ km s$^{-1}$ using the methods described in Grundstrom (2007), and had a range of $\pm 5$ km s$^{-1}$. We then examined the Fourier transforms of the radial velocities with the {\tt Period04} routine (Lenz \& Breger 2005). Our time-series analysis yielded a significant peak with $f = 0.00587 \pm 0.00014 $ cycles d$^{-1}$, or a period of $P=170.36 \pm 4.27$ d. This peak had a signal-to-noise ratio of $\approx 4.5$, implying that this period is significant to 4.5$\sigma$. The phased radial velocity curve is shown in Figure \ref{fig1}. We held the period constant and fit a single-lined orbit. An elliptical and a circular fit yielded indistinguishable results when using the probability test for elliptical solutions of Lucy (2005), so we adopt the circular orbit shown in Fig.~2 with orbital elements given in Table 2. Note that the scatter about the fit is about the same as the amplitude of the fit, which is larger than the individual errors of each radial velocity point. This increased dispersion is likely caused by randomly varying small-scale structures in the disk (e.g., as the result  of non-asymmetric mass ejections from the central star). To reduce this dispersion, we averaged the points in phase bins, and calculated the error of the mean for each phase bin, as shown in the right panel of Fig.~2. These results reveal a statistically significant amplitude of $K_1 = 2.3\pm0.4$ km s$^{-1}$, which corresponds to a 5$\sigma$ detection. If we phase the velocities on other potential periods, such as the 182.8 d period found by Folsom et al.~(2016), the sinusoidal pattern is not as obvious.


We also attempted to use the techniques used previously for the Be stars known to have hot subdwarfs in their UV spectra (Gies et al.~1998, Peters et al. 2008, 2013, 2016). Unfortunately, only four spectra exist in the {\it IUE} archive for \bc, which makes direct detection of the newly found companion unlikely for the given dataset. Nevertheless, we constructed cross-correlation functions between the observed spectra and the TLUSTY models (Lanz \& Hubeny 2007), and shifted these with respect to the expected mass ratio of the system. We performed this in both the entire UV range (1150--1900\AA) and in a short-wavelength range (1150--1200\AA) optimized for these searches. No positive detection was made, primarily due to the inherent noise of the {\it IUE} data and the limited data set. Such detection will likely require the use of {\it HST}/STIS for a direct detection in the future.

\section{Searching for Disk Structure related to the Companion}

Folsom et al.~(2016) reported on H$\alpha$ spectroscopy of \bc\ and measured $V/R$ variations as a function of time. Their results indicated a periodicity of the $V/R$ variations with a period of 183 d, similar to our measurement of the orbital period ($P=170.36$ d). Since current modeling of Be binaries shows disks with density waves driven by the tidal influence of the secondary (Panoglou et al.~2016), it is worth looking for structure in the H$\alpha$ $V/R$ variations as a function of orbital phase. Since the violet ($V$) and red ($R$) peak heights sample the density in the approaching and receding portions of the disk, respectively, we should expect to see a periodic $V/R$ variation as the tidally induced density wave moves around the disk synchronously with the binary.

\begin{figure} 
\begin{center} 
\includegraphics[angle=90, width=8cm]{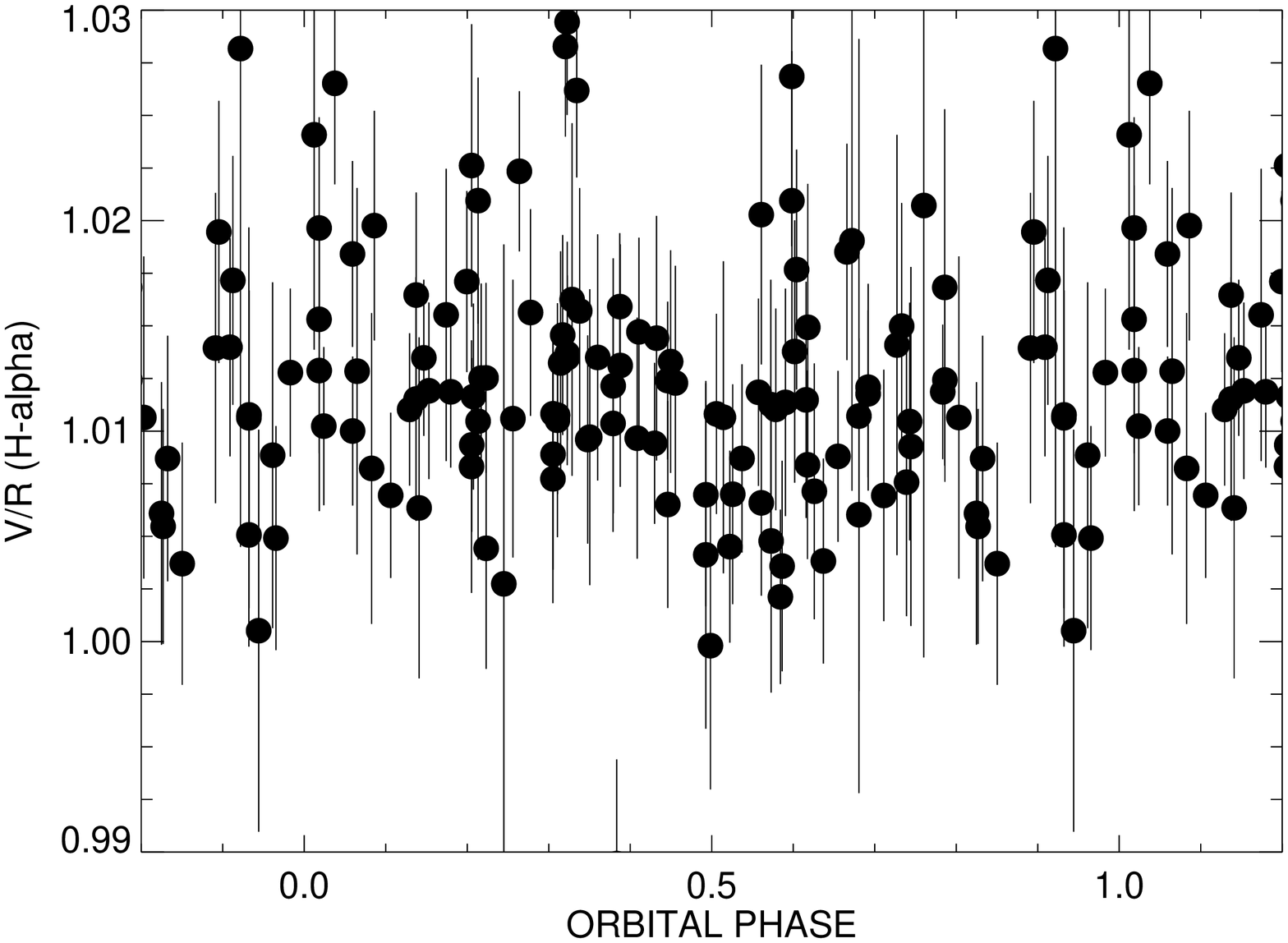}
\includegraphics[angle=90, width=8cm]{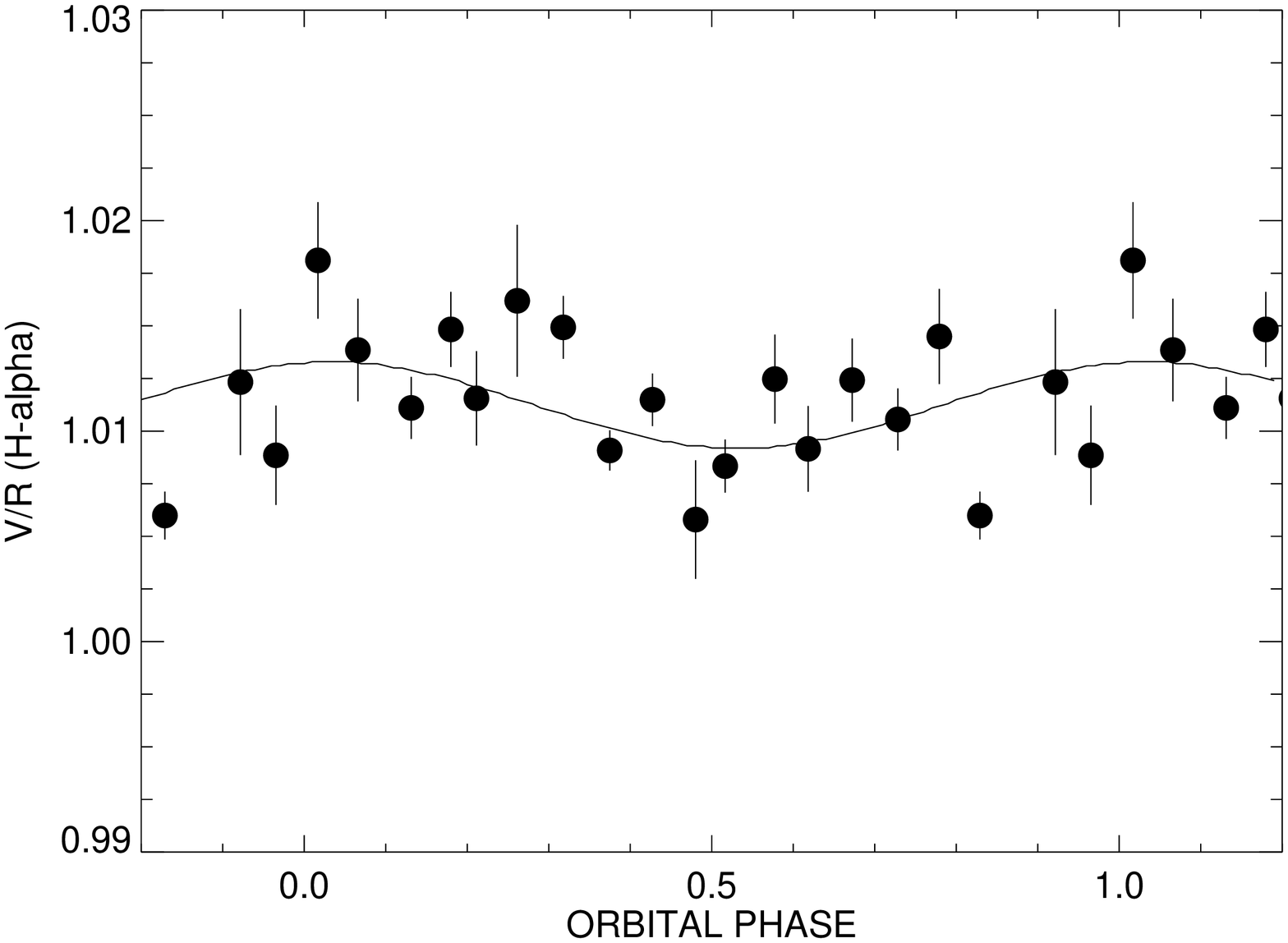}

\end{center} 
\caption{$V/R$ measurements as a function of orbital phase for all H$\alpha$ measurements (left), and phase binned (right).} 
\label{fig2} 
\end{figure}

In Fig.~\ref{fig2}, we plot the $V/R$ ratio as a function of orbital phase for the optical H$\alpha$ line obtained with Ritter Observatory. These measurements are included in Table 1. The $V/R$ measurements were made by fitting two Gaussians to the double-peaked emission, with no removal of the continuum. Error bars were estimated by adding in quadrature the contributions of the noise in the continuum placement and the errors on the fits of Gaussians. We did not subtract the photospheric contribution from the $V$ or $R$ peak height in the calculation of $V/R$.

An analysis of the $V/R$ variations, weighted with the corresponding error bars, of H$\alpha$ with the {\tt Period04} software (Lenz \& Breger 2005) revealed the strongest frequency corresponded to a period of 347 d, or roughly twice the orbital period. A secondary frequency near the orbital frequency is marginally detected. Upon phase-binning the data in the same manner as the orbital analysis (\S 3), we find that there is some moderate variability present at the $\sim1\%$ level. Namely, the $V/R$ ratio mimics the radial velocity (peak near phase 0, minimum near phase 0.5). We fit this binned $V/R$ curve with a simple sine wave forced to the orbital period, to obtain an amplitude of $0.0020\pm0.0003$, and that the maximum is at phase $\phi=0.03\pm 0.03$, verifying the visual estimates of the variability.

\section{Discussion}

The cause of rapid rotation in the classical Be stars is still a debated issue. McSwain \& Gies (2005) show that most Be stars are somewhat evolved and that the frequency increases with age, although there are some young Be stars near the zero-age main-sequence stars. In order to reconcile the ages and spin rates of Be stars, some authors have discussed a redistribution of internal angular momentum as the stars conclude their core H-burning that spins up the outer layers of the star as the core contracts (e.g., Ekstr\"om et al.~2008). Another possibility is that Be stars are the product of binary interaction processes, which likely dominates the evolution of massive stars (Sana et al.~2012, de Mink et al.~2013). 

In the case of \bc\ the binary evolution channel is a likely mechanism for understanding the evolution of the system and the critical rotation rate of the primary B8Ve star (Klement et al.~2015). Vanbeveren et al.~(1998) describe a probable evolutionary channel for the early Be star $\phi$ Per, which was discovered to have a hot subdwarf companion by Gies et al.~(1998) using ultraviolet spectroscopy. Recently, the companion has also been detected directly through interferometry (Mourard et al.~2015). In the case of $\phi$ Per, Vanbeveren et al.~(1998) describe how a system that began its life as a $6+5 M_\odot$ with an orbital period of 13.5 d undergoes conservative Roche lobe overflow. In this situation, the more massive $6 M_\odot$ star loses mass to the mass gainer (the $5M_\odot$ star) until the star has only $\sim 1.5 M_\odot$ of material. Then, as the Roche lobe overflow ends, the secondary star has $9.5 M_\odot$ of material and is rotating at a critical rate. This star becomes the dominant light source and a Be star with a small, $1.5M_\odot$ companion with a long orbital period of 126 d. 
This situation appears to be common for a wide variety of Be stars. Similar direct ultraviolet detections of hot subdwarf companions have  been accomplished for FY CMa (Peters et al.~2008), 59 Cyg (Peters et al.~2013), HR 2142 (Peters et al.~2016), and $o$ Pup (Rivinius et al.~2012, Koubsk\'y et al.~2012) all with small mass functions in the range of $2\times10^{-3} - 1.3 \times 10^{-2} M_\odot$.  Other Be binaries that have not had direct detections of the companion stars such as $\gamma$ Cas (Smith et al.~2012) or $\zeta$ Tau (Ru\v{z}djak et al.~2009) also show similar mass functions ($f(M) = 1 \times 10^{-3}$ and $f(M) = 6\times 10^{-3}$ respectively) that indicate low-mass companions. 

The mass function of the single-lined orbit we derived for \bc\ ($f(M) = 2.0\times10^{-3} M_\odot$) is indicative of a very low mass companion, similar to the aforementioned Be binaries with ultraviolet detections of the companions. 
Matson et al.~(2015) investigated the binary system KOI-81 to find that the primary B8V star was orbiting a hot subdwarf companion that may have  a similar evolutionary history as \bc. If we assume the primary has a similar mass to the primary of KOI-81 ($M = 3M_\odot$; Matson et al.~2015) and that the orbit is coplanar with the Be disk that has been resolved with interferometry and modeled by Klement et al.~(2015) to have an inclination of $i=43^{+3^\circ}_{-2^\circ}$, then the mass function yields an estimate of the secondary mass of $M_2 \sim 0.42M_\odot$. If we adopt the primary mass of $M=3.5 M_\odot$ as presented in Klement et al.~(2015), the secondary mass is slightly larger with $M_2 \sim 0.46M_\odot$. Such a system bears remarkable similarity to the white dwarf companion discovered around the B9V star Regulus (Gies et al.~2008).

Klement et al.~(2015) found the disk of \bc\ to be truncated at $35 R_{\rm eq} = 146 R_\odot$. At that distance, a particle in orbit around the $3-3.5 M_\odot$ star would have an orbital period of $109-118$ d, which is close to the 3:2 resonance with our derived orbital period of 170.4 d. This result is, therefore, in rough agreement with the predictions of Klement et al.~(2015) for the location of the truncation radius. The binary companion thus provides not only a mechanism for the truncation of the disk, but also a simple explanation for the location of the truncation. At this time, there are not enough disks shown to have truncation and resonances with the companion to see what resonances are preferred for such a mechanism, but Klement et al.~(2016, submitted) found evidence of disk truncation in 5 of 6 cases showing that unseen binary companions may be quite common.


Lastly, we return to the models of Be systems with companions in coplanar orbits to the Be disk (Panoglou et al.~2016). The images (Fig.~4 of Panoglou et al.~2016) produced in these models suggest the largest density perturbation occurs at the sub-companion point in the disk. Our $V/R$ measurements (Fig.~3) show that indeed the H$\alpha$ line shows weak variations on the orbital time scale, with the radial velocities being synchronized with the $V/R$ variations (compare Fig.~2b and Fig.~3b). When the primary star approaches us ($\phi=0.5$), the spiral arm on the side of the disk facing the companion would also be approaching us. If this enhancement causes increased emission, then the red peak would increase, causing a small decrease in $V/R$, as seen in Fig.~3b. Conversely, at phase $\phi=0$, there is a corresponding increase in $V/R$. Further observations are necessary to confirm this relationship, although several other binary Be stars show orbital phase-modulated $V/R$ variations (e.g., $\phi$ Per; Gies et al.~1998). 

\section{Conclusions}

This study into the disk and properties into the late-type Be star \bc\ has found the following properties:
\begin{itemize}
\item \bc\ is a binary star system, with an orbital period of 170 d, and an amplitude of 2.25 km s$^{-1}$. The derived mass function is indicative of a small companion.
\item The $V/R$ measurements of the H$\alpha$ profile show that the $V/R$ behavior correlates with the orbit, showing that the outer disk structure is driven by the companion star.
\item Lastly, the properties of the system shows that the system is a strong candidate for past Roche lobe overflow in the system causing a mass reversal and spinning up the now-primary star. 
\end{itemize}

Future studies of \bc\ should focus on improving the orbit through analysis of photospheric lines. Likely, this means collecting high-resolution blue spectroscopy to sample lines such as \ion{He}{1} or \ion{Mg}{2}. The largest difficulty is obtaining high precision velocities with the rapid rotation, but this has been accomplished for the rapid rotator Regulus (Gies et al.~2008). Follow-up spectroscopy in the ultraviolet could easily show the spectroscopic signs of the companion, as evidenced by the results of Matson et al.~(2015), Peters et al.~(2008, 2013, 2016), or Gies et al.~(1998). Lastly, emission lines that form at very large radii should be investigated to look for further structure caused in the presence of a companion star. The long-term stability of the Be disk in the presence of the companion star should allow for new constraints and insights into the Be phenomenon.




\acknowledgments

We thank our referee, Helmut Abt, for a careful review of the manuscript.
We also thank Douglas R. Gies (Georgia State University) for constructive comments that led to an improved presentation of this paper. This research has made extensive use of the Ritter Observatory one-meter telescope, and we are grateful for the continuing support of the University of Toledo and the telescope technician, Michael Brown. Many observations presented in this paper were taken by past members of the Ritter Observing Team, including the following people: Anatoly Miroshnichenko, David Knauth, William Fischer, John Wisniewski, Joshua Thomas, Amanda Gault, Douglas Long, Erica Hesselbach, Greg Thompson, Adam Ritchey, and James Davidson.

ND acknowledges support of the NSF REU program at the University of Toledo under grant No.~1262810.
NDR is grateful for postdoctoral support by the University of Toledo and by the Helen Luedke Brooks Endowed Professorship.
JEB, KSB, and ADB acknowledge support of the NSF through grant AST-1412135.
The research of R.~K. was supported by grant project number 1808214 of the Charles University Grant Agency (GA UK).
A.C.C. acknowledges support from CNPq (grant 307594/2015-7) and FAPESP (grant 2015/17967-7). 

{\it Facilities:} 
\facility{Ritter Observatory (\'echelle spectrograph)}

\clearpage

\begin{deluxetable}{l c c c c } 
\tablecaption{H$\alpha$ measurements. Table 1 is published in its entirety online in the machine-readable format. }
\tablewidth{0pc} 
\centering
\tablehead{ 
\colhead{Date}                      & 
\colhead{RV}    &
\colhead{$\sigma$(RV)} &
\colhead{$V/R$} &
\colhead{$\sigma$($V/R$)} \\
\colhead{(HJD$-$2,450,000)}     & 
\colhead{(km s$^{-1}$)}    &
\colhead{(km s$^{-1}$)} &
\colhead{} &
\colhead{}  } 
\startdata 
1571.7069  & \nodata & \nodata &  1.013  &  0.011\\ 
1586.6321  &  0.90  &  0.63  &  1.009  &  0.009\\ 
1606.6968  &  0.34  &  0.61  &  1.009  &  0.008\\ 
1608.5705  &  -2.70  &  0.77  &  1.019  &  0.010\\ 
1609.6186  &  1.32  &  2.39  &  1.019  &  0.024\\ 
1887.8216  &  -5.41  &  0.83  &  1.008  &  0.012\\ 
1887.8435  &  -1.21  &  0.76  &  1.009  &  0.011\\ 
1919.7665  &  1.13  &  1.54  &  1.004  &  0.017\\ 
1919.7890  &  3.31  &  0.90  &  1.007  &  0.011\\ 
1920.7410  &  3.65  &  1.29  &  1.000  &  0.014\\ 
1935.7301  &  2.07  &  0.75  &  1.004  &  0.010\\ 
1937.7435  &  -3.70  &  1.47  &  1.027  &  0.016\\ 
1937.7603  & \nodata & \nodata &  1.021  &  0.014\\ 
1951.7700  &  2.69  &  1.53  &  1.006  &  0.017\\ 
1951.7856  & \nodata & \nodata &  1.011  &  0.036\\ 
1959.7331  &  2.25  &  1.57  &  1.014  &  0.020\\ 
1961.7362  &  3.83  &  1.09  &  1.008  &  0.013\\ 
1962.6453  &  3.62  &  1.44  &  1.009  &  0.017\\ 
1969.7207  &  -2.45  &  1.58  &  1.017  &  0.017\\ 
1969.7539  &  2.97  &  0.70  &  1.012  &  0.010\\ 
1976.7007  &  0.31  &  0.92  &  1.005  &  0.011\\ 
1980.6886  &  4.80  &  0.81  &  1.004  &  0.012\\ 
1990.6681  &  9.35  &  0.73  &  1.014  &  0.010\\ 
1994.6202  &  4.49  &  1.55  &  1.011  &  0.018\\ 
1994.6426  &  4.14  &  0.66  &  1.005  &  0.011\\ 
1996.6668  &  4.65  &  1.86  &  1.001  &  0.019\\ 
2586.8700  &  -0.08  &  0.96  &  1.010  &  0.011\\ 
2612.8052  &  -0.47  &  0.69  &  1.007  &  0.009\\ 
2615.8279  &  -0.23  &  0.79  &  1.011  &  0.010\\ 
2646.7818  & \nodata & \nodata &  1.021  &  0.043\\ 
2657.7849  &  4.09  &  1.24  &  1.006  &  0.012\\ 
2669.7763  &  0.81  &  1.18  &  1.019  &  0.012\\ 
2681.7018  &  4.39  &  0.95  &  1.005  &  0.011\\ 
2684.7235  &  -0.03  &  0.61  &  1.013  &  0.008\\ 
2690.7221  &  3.16  &  1.30  &  1.015  &  0.013\\ 
2690.7533  &  2.51  &  1.35  &  1.013  &  0.013\\ 
2691.7129  & \nodata & \nodata &  1.010  &  0.008\\ 
2697.7325  &  7.02  &  0.52  &  1.010  &  0.007\\ 
2698.6753  &  6.12  &  1.59  &  1.013  &  0.017\\ 
2701.7080  &  2.46  &  1.43  &  1.008  &  0.015\\ 
2705.6891  &  4.08  &  0.57  &  1.007  &  0.008\\ 
2709.6560  &  6.94  &  0.53  &  1.011  &  0.007\\ 
2712.6170  &  -0.20  &  0.50  &  1.013  &  0.007\\ 
2713.6823  &  -0.59  &  0.62  &  1.012  &  0.008\\ 
2722.6408  &  3.77  &  0.59  &  1.009  &  0.008\\ 
2739.6247  &  4.12  &  0.50  &  1.011  &  0.007\\ 
2740.5860  &  3.64  &  0.50  &  1.011  &  0.007\\ 
2741.6485  &  2.58  &  0.83  &  1.015  &  0.009\\ 
2742.5987  &  0.20  &  0.71  &  1.014  &  0.011\\ 
2742.6199  &  3.20  &  0.60  &  1.013  &  0.008\\ 
2975.8496  &  -1.88  &  0.66  &  1.012  &  0.010\\ 
2743.6133  & \nodata & \nodata &  1.016  &  0.017\\ 
2753.5878  &  2.24  &  0.49  &  1.016  &  0.007\\ 
2757.5798  & \nodata & \nodata &  1.015  &  0.009\\ 
2763.5765  &  -1.72  &  0.47  &  1.012  &  0.007\\ 
2952.9281  &  2.81  &  0.61  &  1.012  &  0.009\\ 
2975.8496  &  -3.41  &  0.50  &  1.012  &  0.008\\ 
3051.7103  &  6.11  &  0.90  &  1.012  &  0.012\\ 
3051.7311  &  5.24  &  0.70  &  1.016  &  0.010\\ 
3063.6942  &  4.04  &  0.71  &  1.012  &  0.009\\ 
3064.6850  &  -0.21  &  1.16  &  1.010  &  0.013\\ 
3075.6373  &  2.59  &  0.85  &  1.016  &  0.010\\ 
3087.5933  &  -3.35  &  0.83  &  1.010  &  0.010\\ 
3101.5654  &  -4.59  &  0.58  &  1.009  &  0.008\\ 
3319.8357  &  1.72  &  1.08  &  1.007  &  0.012\\ 
3398.7043  &  0.44  &  1.31  &  1.016  &  0.014\\ 
3412.6736  &  -2.37  &  1.14  &  1.011  &  0.013\\ 
3426.6358  &  -3.25  &  0.94  &  1.016  &  0.012\\ 
3428.7297  &  -1.47  &  1.29  &  1.010  &  0.014\\ 
3433.5988  &  -2.84  &  0.81  &  1.010  &  0.010\\ 
3433.6294  &  -0.22  &  1.19  &  1.012  &  0.012\\ 
3442.6953  &  0.20  &  1.05  &  1.014  &  0.012\\ 
3446.6330  &  8.22  &  0.91  &  1.012  &  0.011\\ 
3456.6685  &  -4.56  &  1.31  &  1.011  &  0.015\\ 
3458.6149  &  5.66  &  0.82  &  1.007  &  0.010\\ 
3464.5823  &  -1.95  &  1.12  &  1.020  &  0.014\\ 
3466.5754  &  5.81  &  0.96  &  1.011  &  0.012\\ 
3469.6358  &  -0.62  &  0.87  &  1.011  &  0.011\\ 
3471.5645  &  -1.89  &  1.10  &  1.014  &  0.012\\ 
3477.5644  &  2.37  &  0.75  &  1.004  &  0.010\\ 
3495.6049  &  -3.18  &  1.01  &  1.010  &  0.011\\ 
3712.8752  &  6.67  &  0.88  &  1.020  &  0.011\\ 
3744.7318  &  -0.54  &  0.96  &  1.008  &  0.012\\ 
3762.6885  &  -1.34  &  0.84  &  1.011  &  0.011\\ 
3785.7587  &  -2.46  &  0.77  &  1.007  &  0.010\\ 
3798.7281  &  -1.87  &  0.68  &  1.005  &  0.009\\ 
3812.6785  &  -1.06  &  0.83  &  1.018  &  0.011\\ 
3814.6635  &  -1.38  &  0.95  &  1.011  &  0.011\\ 
3834.6419  &  -1.14  &  0.84  &  1.015  &  0.012\\ 
3851.5665  &  0.85  &  0.78  &  1.009  &  0.012\\ 
4038.9421  &  0.76  &  1.14  &  1.011  &  0.012\\ 
4111.8313  &  2.00  &  1.08  &  1.014  &  0.012\\ 
4136.6385  &  -3.71  &  0.78  &  1.011  &  0.010\\ 
4155.6664  &  -0.76  &  0.68  &  1.008  &  0.009\\ 
4432.9101  & \nodata & \nodata &  1.003  &  0.032\\ 
4488.7881  &  -2.30  &  1.19  &  1.005  &  0.014\\ 
4546.6284  &  0.78  &  1.04  &  1.017  &  0.012\\ 
4563.6150  &  -0.29  &  2.00  &  1.024  &  0.020\\ 
4571.6351  &  0.60  &  0.74  &  1.018  &  0.009\\ 
4769.9063  &  3.28  &  0.83  &  1.013  &  0.009\\ 
4769.9396  &  -2.09  &  1.02  &  1.004  &  0.011\\ 
4797.9489  &  -2.66  &  1.12  &  1.013  &  0.012\\ 
4868.7383  &  -5.28  &  1.51  &  1.011  &  0.015\\ 
4883.7126  &  -1.34  &  1.46  &  1.014  &  0.015\\ 
4895.6562  &  -2.25  &  1.63  &  1.009  &  0.016\\ 
4908.6496  &  -3.86  &  0.79  &  1.027  &  0.010\\ 
4938.6058  &  -1.76  &  1.04  &  1.021  &  0.012\\ 
5137.9482  &  -3.79  &  0.92  &  0.989  &  0.010\\ 
5177.8790  &  -3.41  &  1.27  &  1.015  &  0.014\\ 
5229.6448  &  0.70  &  4.33  &  1.028  &  0.047\\ 
5257.6279  &  -0.41  &  0.90  &  1.020  &  0.011\\ 
5273.6010  &  -4.59  &  0.49  &  1.012  &  0.007\\ 
5296.5849  &  -4.98  &  0.89  &  1.013  &  0.011\\ 
5297.5668  &  -4.07  &  0.52  &  1.028  &  0.009\\ 
5512.8658  &  -4.71  &  0.59  &  1.002  &  0.008\\ 
5607.7060  &  -4.36  &  1.46  &  1.006  &  0.016\\ 
7393.8888  &  0.19  &  1.07  &  1.007  &  0.012\\ 
7420.7281  &  -0.26  &  0.40  &  1.012  &  0.006\\ 
7491.6034  &  -0.46  &  0.71  &  1.017  &  0.009\\ 
7492.6005  &  -1.04  &  1.25  &  1.023  &  0.013\\ 
7494.5920  &  -0.94  &  0.76  &  1.013  &  0.009\\ 
7502.5815  &  0.09  &  0.52  &  1.022  &  0.008\\ 
7512.5694  &  1.26  &  0.66  &  1.029  &  0.009\\ 
7514.5672  &  -1.66  &  0.61  &  1.026  &  0.008\\\enddata
\end{deluxetable}

\begin{deluxetable}{lc} 
\tablewidth{0pc} 
\tablecaption{Circular Orbital Elements\label{orbit}} 
\tablehead{ 
\colhead{Element}                      & \colhead{Value}       } 
\startdata 
$P$ (d)                       \dotfill & $170.4\pm4.3$ \\ 
$T_{\rm RV max}$ (HJD -- 2,450,000)        \dotfill & $ 2001.2 \pm  5.0$     \\ 
$K_1$ (km s$^{-1}$)           \dotfill & $2.25 \pm 0.4$         \\ 
$a_1\sin i$ ($R_\odot$)       \dotfill &  $7.6\pm2.3  $          \\ 
$f(M)$ ($M_\odot$)            \dotfill & $(2.0\pm 0.5)\times 10^{-3}$            \\ 
r.m.s. (km s$^{-1}$)          \dotfill &         3.3             \\ 
\enddata 

\end{deluxetable} 

\clearpage


\begin{thebibliography}{}

\bibitem[Abt(1979)]{abt}
	Abt, H., 1979, \apj, 230, 485
\bibitem[Carciofi et al.(2006)]{2006ApJ...652.1617C}
	Carciofi, A. C., Miroshnichenko, A. S., Kusakin, A. V., et al.
	2006, \apj, 652, 1617
\bibitem[Carciofi et al.(2009)]{2009A&A...504..915C}
	Carciofi, A. C., Okazaki, A. T., Le Bouquin, J.-B., et al.
	2009, \aap, 504, 915
\bibitem[de Mink et al.(2013)]{2013ApJ...764..166D}
	de Mink, S. E., Langer, N., Izzard, R. G., Sana, H., \& de Koter, A.
	2013, \apj, 764, 166
\bibitem[Ekstr\"om et al.(2008)]{2008A&A...478..467E}
	Ekstr\"{o}m, S., Meynet, G., Maeder, A., Barblan, F.
	2008, \aap, 478, 467
\bibitem[Folsom et al.(2016)]{folsom}
	Folsom, L., Miroshnichenko,A. S., Danford, S., Zharikov, S. V., \& Sawicki, C. J.,
	2016, in Bright Emissaries: Be Stars as Messengers of Star-Disk Physics, in
	press, ed. A. Sigut \& C. Jones, Astronomical Society of the Pacific Conference
	Series
\bibitem[Gies et al.(1998)]{1998ApJ...493..440G}
	Gies, D. R., Bagnuolo, W. G., Jr., Ferrara, E. C., Kaye, A. B., Thaller, M. L., Penny, L. R., Peters, G. J.
	1998, \apj, 493, 440
\bibitem[Gies et al.(2008)]{2008ApJ...682L.117G}
	Gies, D. R., Dieterich, S., Richardson, N. D., et al.
	2008, \apjl, 682, 117
\bibitem[Grundstrom (2007)]{grundstrom}
	Grundstrom, E. D. 2007, PhD thesis, Georgia State Univ., available at http://scholarworks.gsu.edu/phy\_astr\_diss/19/
\bibitem[Jarad et al.(1989)]{1989MNRAS.238.1085J}
	Jarad, M. M., Hilditch, R. W., \& Skillen, I.
	1989, \mnras, 238, 1085
\bibitem[Klement et al.(2015)]{2015A&A...584A..85K} 
	Klement, R., Carciofi, A. C., Rivinius, Th., et al.
	2015, \aap, 584, 85
\bibitem[Koubsk\'y et al.(2012)]{koubsky}
	Koubsk\'y, P., Kotkov\'a, L., Votruba, V., \v{S}lechta, M., \& Dvo\v{r}\'akov\'a, \v{S}.
	2012, \aap, 545, 121	
\bibitem[Kraus et al.(2012)]{kraus}
	Kraus, S., Monnier, J. D., Che, X., et al.
	2012, \apj, 744, 19
\bibitem[Lanz \& Hubeny(2007)]{tlustybstar}
	Lanz, T., \& Hubeny, I. 
	2007, ApJS, 169, 83
\bibitem[Lee et al.(1991)]{lee}
	Lee, U., Osaki, Y., \& Saio, H.
	1991, \mnras, 250, 432
\bibitem[Lenz \& Breger(2005)]{2005CoAst.146...53L}	
	Lenz P., \& Breger M. 
	2005, CoAst, 146, 53
\bibitem[Lucy (2005)]{2005A&A...439..663L}
	Lucy, L. B.
	2005, \aap, 439, 663	
\bibitem[Matson et al.(2015)]{2015ApJ...806..155M}
	Matson, R. A., Gies, D. R., Guo, Z., Quinn, S. N., Buchhave, L. A., Latham, D. W., Howell, S. B., \& Rowe, J. F.
	2015, \apj, 806, 155
\bibitem[McSwain \& Gies (2005)]{2005ApJS..161..118M}
	McSwain, M. V., \& Gies, D. R.
	2005, \apjs, 161, 118
\bibitem[Mourard et al.(2015)]{2015A&A...577A..51M}
	Mourard, D., Monnier, J. D., Meilland, A., et al.
	2015, \aap, 577, 51
\bibitem[Okazaki et al.(2002)]{2002MNRAS.337..967O}
	Okazaki, A. T., Bate, M. R., Ogilvie, G. I., \& Pringle, J. E.
	2002, \mnras, 337, 967
\bibitem[Panoglou et al.(2016)]{2016MNRAS.461.2616P}
	Panoglou, D., Carciofi, A. C., Vieira, R. G., Cyr, I. H., Jones, C. E., Okazaki, A. T., \& Rivinius, Th.
	2016, \mnras, 461, 2616
\bibitem[Peters et al.(2008)]{2008ApJ...686.1280P} 
	Peters, G. J., Gies, D. R., Grundstrom, E. D., \& McSwain, M. V.
	2008, \apj, 686, 1280
\bibitem[Peters et al.(2013)]{2013ApJ...765....2P} 
	Peters, G. J., Pewett, T. D., Gies, D. R., Touhami, Y. N., \& Grundstrom, E. D.
	2013, \apj, 765, 2
\bibitem[Peters et al.(2016)]{2016arXiv160701829P} 
	Peters, G. J., Wang, L., Gies, D. R., \& Grundstrom, E. D.
	2016, \apj, in press. arXiv: 1607.01829
\bibitem[Polidan (1976)]{polidan}
	Polidan, R. S. 1976, in Be and Shell Stars, ed. A. Slettebak, IAU Symp., 70, 401
\bibitem[Rivinius et al.(2012)]{rivinius}
	Rivinius, Th., Vanzi, L., Chacon, J., et al.
	2012, in Circumstellar Dynamics at High Resolution, eds. A. C. Carciofi, \& T. Rivinius, ASP Conf. Ser., 464, 75
\bibitem[Rivinius, Carciofi, \& Martayan (2013)]{2013A&ARv..21...69R}
	Rivinius, Th., Carciofi, A. C., \& Martayan, C.
	2013, \araa, 21, 69
\bibitem[Ru\v{z}djak et al.(2009)]{2009A&A...506.1319R}
	Ru\v{z}djak, D., Bo\v{z}i\'c, H., Harmanec, P., et al.
	2009, \aap, 506, 1319
\bibitem[Saio(1991)]{saio}
	Saio, H.
	1991, \mnras, 250, 432
\bibitem[Sana et al.(2012)]{sana}
	Sana, H., de Mink, S. E., de Koter, A., et al.
	2012, Science, 337, 444
\bibitem[Schaefer et al.(2010)]{2010AJ....140.1838S}	
	Schaefer, G. H., Gies, D. R., Monnier, J. D., et al.
	2010, \aj, 140, 1838
\bibitem[Shafter et al.(1986)]{1986ApJ...308..765S}
	Shafter, A. W., Szkody, P., \& Thorstensen, J. R., 
	1986, \apj, 308, 765
\bibitem[Smith et al.(2012)]{2012A&A...540A..53S}
	Smith, M. A., Lopes de Oliveira, R., Motch, C., et al.
	2012, \aap, 540, 53
\bibitem[Vanbeveren et al.(1998)]{1998A&ARv...9...63V}
	Vanbeveren, D., De Loore, C., \& Van Rensbergen, W.
	1998, \araa, 9, 63
\bibitem[Wheelwright et al.(2012)]{wheelwright}
	Wheelwright, H. E., Bjorkman, J. E., Oudmaijer, R. D., Carciofi, A. C., Bjorkman, K. S., \& Porter, J. M.	
	2012, \mnras, 423, L11
\end{thebibliography}
\end{document}